\documentclass[sigconf,authorversion]{acmart}
\AtBeginDocument{%
  }

\copyrightyear{2026}
\acmYear{2026}
\setcopyright{cc}
\setcctype{by}
\acmConference[VRST '26]{32nd ACM Symposium on Virtual Reality Software and Technology}{November 16--18, 2026}{Sendai, Japan}
\acmBooktitle{32nd ACM Symposium on Virtual Reality Software and Technology (VRST '26), November 16--18, 2026, Sendai, Japan}
\acmDOI{10.1145/3822517.3848655}
\acmISBN{979-8-4007-2811-2/2026/11}

\usepackage{tabu}                      %
\usepackage{booktabs}                  %
\usepackage{lipsum}                    %
\usepackage{mwe}                       %
\usepackage{amsmath}  
\usepackage{mathptmx}                  %
\usepackage{svg}
\usepackage{graphicx}    %
\usepackage{subcaption}
\usepackage{placeins}

\usepackage{color}
\usepackage{soul}

\begin{document}

\title{Comparing Haptic Feedback Across Hand Tracking and Controllers in VR Object Interaction Tasks}

\author{Natalia Ocampo}
\affiliation{%
  \institution{Carleton University}
  \city{Ottawa}
  \country{Canada}
}
\email{nataliaocampo@carleton.ca}
\orcid{0009-0002-5929-9927}

\author{J. Felipe Gonzalez}
\affiliation{%
  \institution{Carleton University}
  \city{Ottawa}
  \country{Canada}
  }
\email{johannavila@carleton.ca}
\orcid{0000-0002-0716-1689}

\author{Robert J. Teather}
\affiliation{%
  \institution{Monash University}
  \city{Melbourne}
  \country{Australia}
}
\email{rob.teather@monash.edu}
\orcid{0009-0007-4572-1820}

\author{Kiyoshi Kiyokawa}
\affiliation{%
  \institution{NAIST}
  \city{Ikoma}
  \country{Japan}
}
\email{kiyo@is.naist.jp}
\orcid{0000-0003-2260-1707}

\renewcommand{\shortauthors}{Ocampo et al.}

\begin{abstract}
  Hand tracking offers natural VR interaction but lacks controllers' inherent tactile feedback, while haptic feedback across input methods remains understudied. We conducted a participant study comparing vibration, impulse, and no feedback in hand- and controller-based VR interaction across grasp-, pinch-, and tap-like gestures assessing performance, workload, and preference. A custom glove and controller-mounted device provided both feedback types, aiming for consistency across input modalities. Controllers were faster for grasp and pinch, whereas hand tracking was more accurate for pinch and preferred overall. Haptics had limited performance effects, although impulse reduced grasp accuracy with controllers. Participants preferred vibration and impulse over no haptic feedback, favouring vibration overall. Our findings reveal a more nuanced relationship between haptic feedback, input device, and interaction context than suggested by previous work.

\end{abstract}



\graphicspath{{figs/}{figures/}{pictures/}{images/}{./}}

\maketitle
\section{Introduction}

Modern virtual reality (VR) devices support interaction primarily through handheld controllers, with optical hand tracking becoming increasingly common \cite{metaMetaQuestMR2025a, htcVIVECanadaVIVE2025, sonyPlayStationVRLive2025, PICO4UltraVRMRheadsetPICO}. 
Controllers generally support faster and more accurate interaction, as well as a greater sense of control, ease of use, and comfort across several VR interaction tasks \cite{kangasTradeOffTaskAccuracy2022, luongControllersBareHands2023, masurovskyControllerFreeHandTracking2020}. 
Hand tracking, however, obviates the need for a handheld device and provides more natural mappings between real and virtual hand movements and is often preferred by users \cite{buckinghamHandTrackingImmersive2021a, argelaguetRoleInteractionVirtual2016, adkinsEvaluatingGraspingVisualizations2021, hameedEvaluatingHandtrackingInteraction2021}. 
These input paradigms thus present a recurring design trade-off: controllers tend to better support efficient and precise interaction, while hand tracking offers naturalness, realism, and improved embodiment \cite{kangasTradeOffTaskAccuracy2022, luongControllersBareHands2023, masurovskyControllerFreeHandTracking2020, argelaguetRoleInteractionVirtual2016, adkinsEvaluatingGraspingVisualizations2021, hameedEvaluatingHandtrackingInteraction2021}.

User experience and task effectiveness in VR depend on several factors beyond the input device. %
A key difference between hand tracking and controllers 
is their ability to provide 
tactile feedback. In addition to providing a stable handheld tangible object, current commercial VR controllers provide built-in vibrotactile feedback \cite{metaMetaQuestMR2025a, htcVIVECanadaVIVE2025} during contact events that frequently occur in object acquisition tasks \cite{weeHapticInterfacesVirtual2021, masurovskyControllerFreeHandTracking2020}. Hand tracking, in contrast, usually provides no tactile confirmation \cite{buckinghamHandTrackingImmersive2021a, weeHapticInterfacesVirtual2021}, and additional haptic cues %
remain difficult to integrate into commercial VR systems.
Prior work has explored tactile support for hand tracking. 
Typically, such solutions require external add-on hardware, including commercially available haptic gloves \cite{sensegloveFindOutOur2023, GlovesG1, bhapticsTactGloveDK2Haptic2025}, research glove systems \cite{shenFluidRealityHighResolution2023, baikHapticGloveUsing2020, yamaguchiHandshakeFeedbackHaptic2023}, fingertip devices \cite{moonEffectsWholeHandInteractions2023}, force or impact devices \cite{yoshimuraSkinPressuretypeGrasping2022, lopesImpactoSimulatingPhysical2015}, and passive haptics or physical props \cite{audaEnablingReusableHaptic2021, marichalHandasapropUsingHand2023}. 
However, these approaches often involve trade-offs in comfort, portability, complexity, and cost, limiting their integration into consumer VR systems \cite{vanwegenOverviewWearableHaptic2023, irigoyenNarrativeReviewHaptic2024, patelWearableHapticFeedback2026}. 
In contrast, controller vibration remains comparatively simple, inexpensive, and widely available. 
It thus remains unclear whether haptic feedback provides comparable benefits across hand-tracked and controller-based interaction \cite{moonHandTrackingVibrotactile2023, moonEffectsWholeHandInteractions2023}.

Despite extensive research on input and haptic feedback, their combined effect remains understudied. 
Prior work has compared controllers and hand tracking~\cite{kangasTradeOffTaskAccuracy2022, luongControllersBareHands2023, masurovskyControllerFreeHandTracking2020} or examined haptic devices and cues \cite{weeHapticInterfacesVirtual2021, vanwegenOverviewWearableHaptic2023, patelWearableHapticFeedback2026}. However, few studies have examined input and haptic feedback together within the same VR interaction setting \cite{moonHandTrackingVibrotactile2023, moonEffectsWholeHandInteractions2023}. Differences in their experimental setups limit direct comparison across input methods, while application-specific tasks may not generalize across VR interactions. %
This gap is more pronounced in gesture-based interaction. 
Common VR gestures such as grasping, pinching, and tapping are often treated as comparable despite their different interaction demands \cite{mutasimPinchClickDwell2021, wagnerFittsLawStudy2023, blagaVRGraspHumanGrasp2025}. 
Gestures can constrain haptic feedback delivery \cite{weeHapticInterfacesVirtual2021}%
, while performance and user experience vary between hand-tracked and controller-based input \cite{ocampoComparingHandController2025}. 
Thus, gesture-specific demands may limit the generalizability of findings on input and haptic feedback.

To address these gaps, we evaluated user performance and experience across combinations of input and haptic feedback. Our research question was: \textbf{\textit{How do input and haptic feedback influence interaction across common VR gestures?}} We conducted a study with 24 participants who completed gesture-based object manipulation tasks in a VR environment using two input modalities: controllers and hand tracking, under three haptic feedback conditions: vibration, impulse, and no feedback. The study included three interaction tasks corresponding to grasping, pinching, and tapping, approximating these gestures with the controller through trigger- or collision-based interactions. This enables comparison of input and haptic effects across three common VR gestures. %

To our knowledge, our study is the first to systematically analyze how input and haptic feedback influence task performance and user experience across common VR gestures with different interaction demands. By examining these factors together, this study provides design-relevant evidence for integrating haptics more effectively into practical VR interaction.

\raggedbottom
\section{Related Work}

\subsection{Input Modalities}

Controllers and hand tracking are the most common VR inputs, with controllers remaining dominant on commercial VR platforms such as Meta Quest and HTC Vive \cite{metaMetaQuestMR2025a, htcVIVECanadaVIVE2025}.
Controllers provide a handheld form factor, precise input, passive tactile support, vibrotactile feedback, and long-standing familiarity from gaming and other interactive systems \cite{novacekOverviewControllersUser2020, buckinghamHandTrackingImmersive2021a}. These characteristics make 
controllers well-suited for efficient object manipulation and target-selection tasks, supporting higher accuracy, stability, and responsiveness than free-hand interaction \cite{kangasTradeOffTaskAccuracy2022, luongControllersBareHands2023, masurovskyControllerFreeHandTracking2020}. 
However, controllers map real hand movements less naturally to virtual actions. Because controller interaction typically relies on buttons rather than direct hand-to-action correspondence, prior work has found that controllers offer lower perceived naturalness and embodiment compared with hand-based interaction \cite{buckinghamHandTrackingImmersive2021a, dewezAvatarFriendly3DManipulation2021}. Controllers can also constrain finger use and introduce ergonomic challenges for some users, particularly those with smaller hands \cite{brownEvaluatingVideoGame2013}. In addition, when paired with virtual hand representations, visual-proprioceptive mismatches can further weaken embodiment and reduce interaction quality \cite{pontonStretchYourReach2024, argelaguetRoleInteractionVirtual2016, ocampoComparingHandController2025}. DeMarbre et al. also report that selection performance declines when neither the hand nor controller is visually represented \cite{demarbre2025effects}.

In contrast, hand tracking enables interaction with virtual objects using hand gestures without a handheld intermediary \cite{buckinghamHandTrackingImmersive2021a}. This direct mapping between real and virtual hand movements has been associated with greater perceived naturalness, realism, and embodiment %
\cite{argelaguetRoleInteractionVirtual2016, adkinsEvaluatingGraspingVisualizations2021, hameedEvaluatingHandtrackingInteraction2021}. However, hand tracking is more susceptible to occlusion and latency, and gesture recognition can be unreliable, yielding lower precision and reliability 
in demanding tasks \cite{kangasTradeOffTaskAccuracy2022, luongControllersBareHands2023}. Prior work has shown that users often prefer controllers for performance-oriented tasks; the absence of tactile confirmation in hand tracking can also contribute to poorer usability and less accurate interaction~\cite{masurovskyControllerFreeHandTracking2020, johnsonExploringHandTracking2023}.

Furthermore, an important distinction between these modalities is their provision of haptic feedback. Controllers provide built-in tactile confirmation through embedded vibration, whereas hand tracking typically requires external haptic augmentation to provide comparable tactile cues \cite{weeHapticInterfacesVirtual2021, masurovskyControllerFreeHandTracking2020}. The same haptic cue may therefore serve different roles when delivered through a tangible handheld controller than when applied directly to the hand. How these differences in haptic feedback manifest across controller- and hand-based input remains understudied.

\subsection{Haptic Feedback}

Touch is critical in real-world object manipulation because it confirms contact, supports action control, and helps users interpret interaction outcomes \cite{hoffmanPhysicallyTouchingVirtual1998, muenderHapticFidelityFramework2022, shiHapticSensingFeedback2024}. Such tactile cues are often reduced or absent in VR, where interaction is primarily mediated through visual feedback %
\cite{hoffmanPhysicallyTouchingVirtual1998, muenderHapticFidelityFramework2022, weeHapticInterfacesVirtual2021}. %
VR systems have explored several haptic approaches, including passive props \cite{lindemanUsableVREmpirical1999, besanconMouseTactileTangible2017, huPneuMultiToolsAutoFoldingMultiShapes2019}, wearable devices \cite{vanwegenOverviewWearableHaptic2023, choiGrabityWearableHaptic2017, moonEffectsWholeHandInteractions2023}, and handheld active feedback devices \cite{benkoNormalTouchTextureTouchHighfidelity2016, choiCLAWMultifunctionalHandheld2018, leeTORCVirtualReality2019, richardMultiVibesWhatIf2023}. 

Passive props provide realistic tactile cues by physically matching virtual objects, but they do not scale well because each must be tracked, aligned, and often matched to its virtual counterpart in properties such as weight \cite{choiGrabityWearableHaptic2017} or shape \cite{gonzalezavilaAdapticShapeChanging2021}. Active haptic systems, instead, use actuated devices to recreate tactile properties such as stiffness \cite{ryuElaStickHandheldVariable2020}, weight and torque \cite{shigeyamaTranscaliburWeightShifting2019, leeTORCVirtualReality2019}, internal mass transfer \cite{saghebSWISHShiftingWeightInterface2019}, or shape \cite{gonzalezavilaAdapticShapeChanging2021}. Together, these approaches show that haptic feedback can modify VR user experience and performance.  Vibrotactile feedback, which forms one focus of this work, is among the most widely studied and commonly used approaches \cite{weeHapticInterfacesVirtual2021, brasenEffectsVibrotactileFeedback2019, gibbsComparisonEffectsHaptic2022, islamVibrotactileFeedbackVirtual2022, richardMultiVibesWhatIf2023, chenInvestigatingDifferentModalities2018, kaulHapticHeadSphericalVibrotactile2017, hendersonLeveragingDistalVibrotactile2019}. %

Vibrotactile cues have been explored across several form factors. For example, head-mounted vibrotactile displays have been used for spatial guidance, helping users locate virtual targets without overloading the visual channel \cite{chenInvestigatingDifferentModalities2018} and, in some cases, improving target-finding speed and precision compared with spatial audio \cite{kaulHapticHeadSphericalVibrotactile2017}. Vibrotactile feedback has also been embedded in tangible and handheld devices to support richer object manipulation. Cabaret et al. \cite{cabaretDoesMultiActuatorVibrotactile2024} showed that users can discriminate localized vibrotactile cues in handheld tangibles, while other systems have combined vibration with impact, shear, skin pressure, or weight cues to simulate contact and tool sensations~\cite{yoshimuraSkinPressuretypeGrasping2022, kimMMGripHandheldMultimodal2022, patelPowVRtoolHandheldHaptic2025}. Wearable and fingertip-based systems have also been used to study vibrotactile feedback in manual VR tasks. Kreimeier et al. \cite{kreimeierEvaluationDifferentTypes2019} found that vibrotactile feedback increased presence compared with visual-only feedback, although force feedback was more effective at reducing task completion time in some tasks. Kourtesis et al. \cite{kourtesisActionSpecificPerceptionPerformance2022} further showed that vibrotactile and electrotactile feedback can affect target-acquisition performance differently, suggesting that tactile feedback modality influences perception and performance \cite{stankeTactileWearComparisonElectrotactile2020, sugaSoftnessPresentationCombining2023}. Overall, vibrotactile feedback is flexible, but its effects depend on its delivery and interaction context.

Related work has explored alternatives beyond vibration, including impulse- and impact-like feedback for contact events. For example, Yoshimura et al. \cite{yoshimuraSkinPressuretypeGrasping2022} developed a device that delivered brief force impulses during virtual racket-ball contact. Impulse improved perceived realism compared with conventional VR controller vibration. Kim et al. \cite{kimMMGripHandheldMultimodal2022} similarly developed a handheld multimodal device combining vibration, impact, and shear cues to render more varied and realistic collision events. Lopes et al. \cite{lopesImpactoSimulatingPhysical2015} combined a solenoid tap with electrical muscle stimulation to render the tactile contact and impulse response of physical impact, producing more realistic sensations than either technique alone. Together, these studies demonstrate that impulse feedback can convey aspects of physical contact that vibration alone may not reproduce.

Closest to our study, Moon et al. \cite{moonHandTrackingVibrotactile2023, moonEffectsWholeHandInteractions2023} examined vibrotactile feedback across controller-based and hand-tracked interactions. Their first study compared controllers, hand tracking, and hand tracking with fingertip vibrotactile gloves in a VR rhythm game \cite{moonHandTrackingVibrotactile2023}. They found that both input and haptic feedback influenced performance and user experience, but the use of a game-based task limited generalization to fundamental interactions such as target acquisition or object manipulation. Moreover, since they compared the built-in vibration of commercial controllers to custom vibration gloves, their study introduced potential hardware confound effects. Their follow-up study \cite{moonEffectsWholeHandInteractions2023} examined controllers and hand tracking with and without single-fingertip vibrotactile feedback in a cooperative VR game, confirming that vibrotactile feedback affects presence and engagement, while controllers were faster.

We build on prior work using a custom haptic system producing both vibrotactile and impulse feedback across controller-based and hand-tracked interactions. Like Moon et al. \cite{moonHandTrackingVibrotactile2023}, we developed a custom glove compatible with the headset's built-in hand tracking. We further adapted the same vibrotactile technology and cable-driven impulse mechanism to the controller, reducing hardware-related differences to facilitate a closer comparison across input modalities.

\begin{figure}[htpb!]
  \centering
  \begin{subfigure}[]{0.45\columnwidth}
    \includegraphics[width=\linewidth]{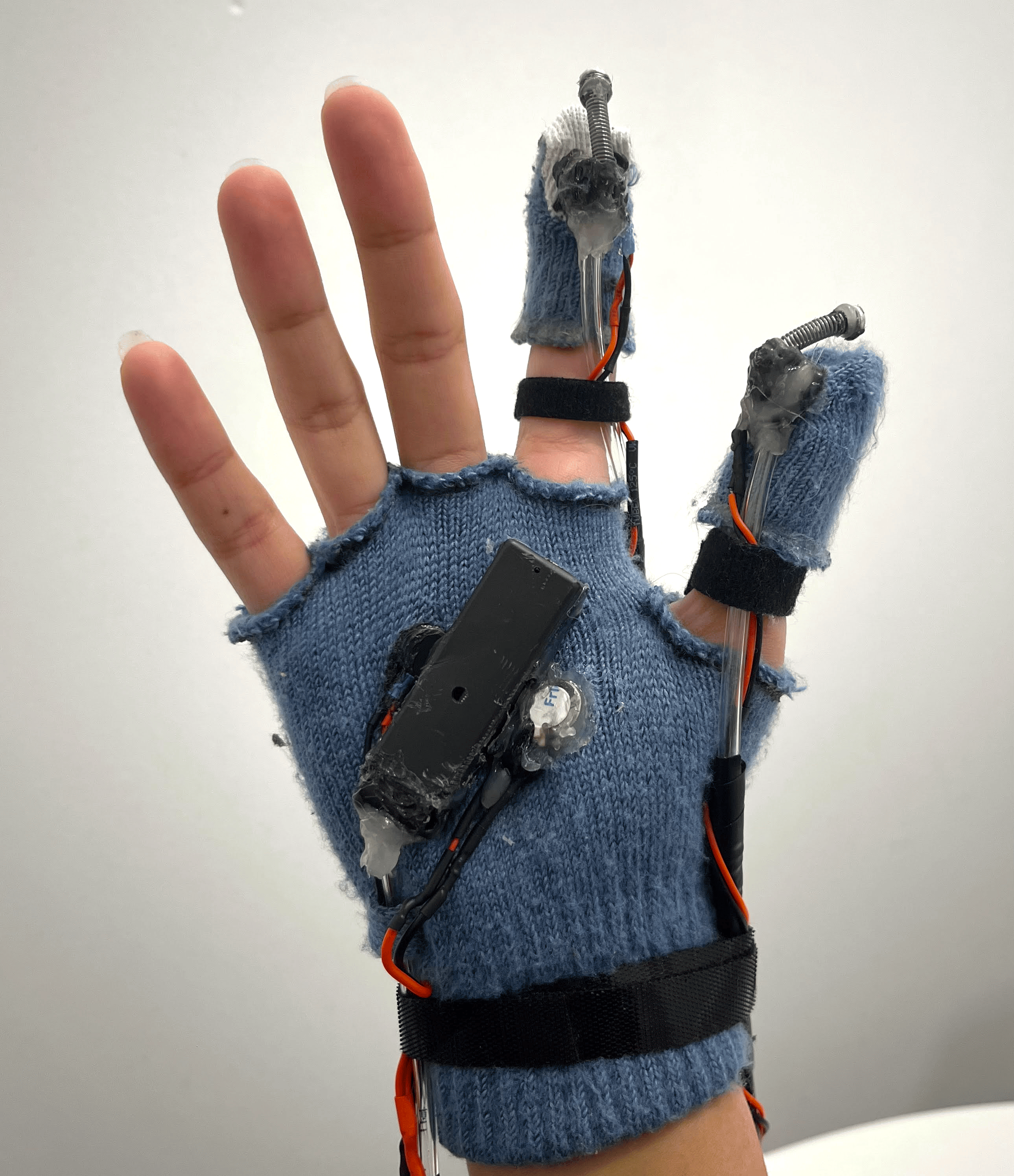}
    \caption{Glove }
    \label{subfiga:haptic-glove}
  \end{subfigure}
  \hspace{1em}
  \begin{subfigure}[]{0.45\columnwidth}
    \includegraphics[width=\linewidth]{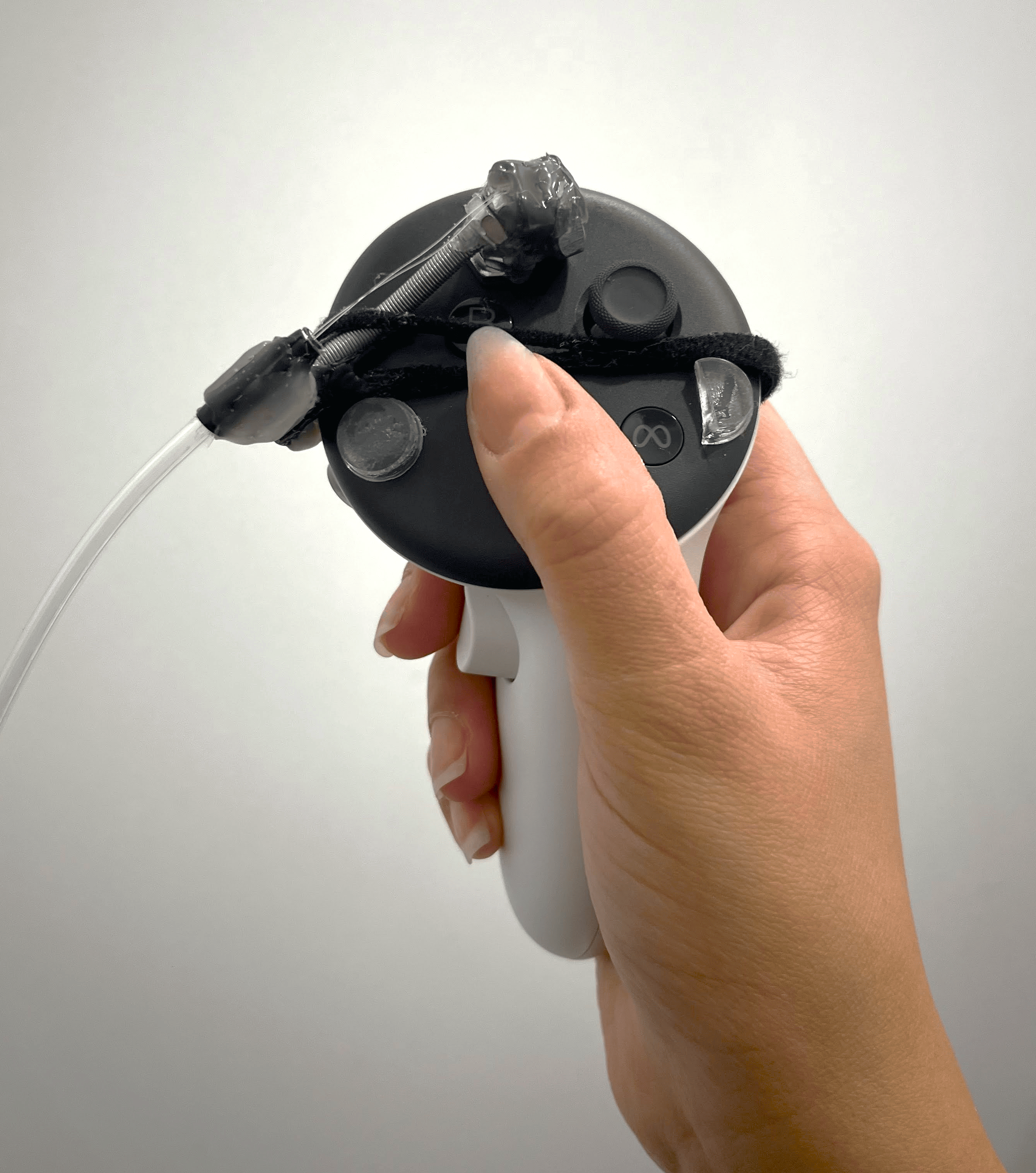}
    \caption{Controller}
    \label{subfigb:haptic_controller}
  \end{subfigure}
  \vspace{-1em}
  \caption{Custom haptic devices that deliver vibration and impulse feedback: (a) haptic glove for hand tracking and (b) controller-mounted device for controller input.}
  \label{fig:haptic-devices}
\end{figure}

\begin{figure*}[htbp!]
  \centering
  \includegraphics[width=1.0\textwidth]{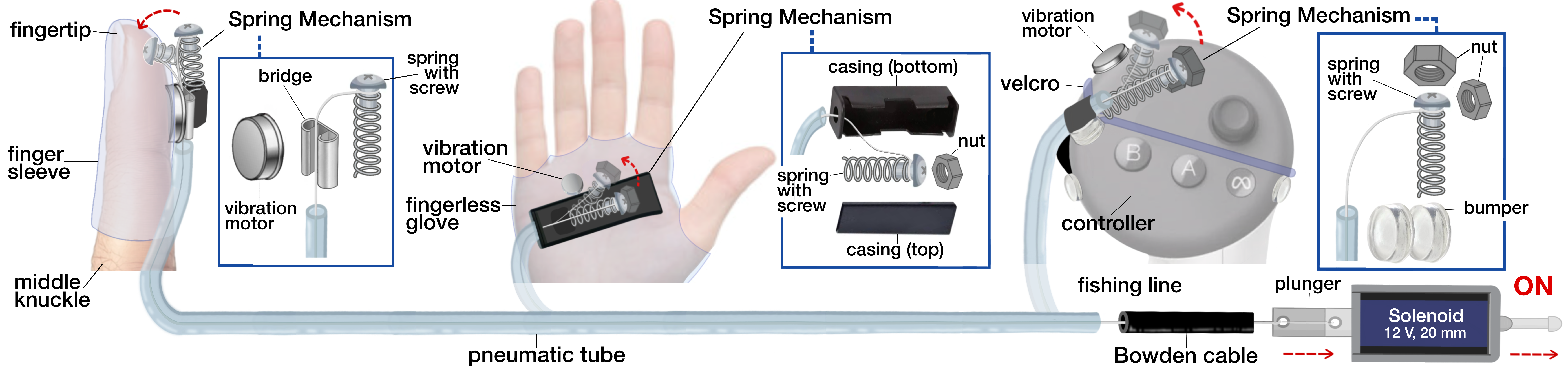}
  \caption{Haptic modules with exploded diagrams: (left) finger-sleeve and (middle) palm modules for the haptic glove, and (right) controller-mounted module for the controller device. Each uses a dedicated solenoid (bottom) to actuate the impulse spring mechanism.}
  \label{fig:haptic-modules}
\end{figure*}

\section{Haptic Device Design}
\label{sec:hapticdevices}

We developed a custom haptic system (left- and right-hand versions) to compare vibration and impulse across hand-tracked and controller-based input. The system included a haptic glove and a controller-mounted device (Figure~\ref{fig:haptic-devices}). Unlike prior work comparing custom wearable feedback with standard controller vibration~\cite{moonEffectsWholeHandInteractions2023}, both devices used the same vibration hardware, cable-driven impulse mechanism, and control logic. This improved hardware consistency across input modalities, although haptic feedback was not precisely equivalent since mounting location and gesture mapping differed.
Both devices were compact, low-cost, and compatible with the Meta Quest 3. The glove weighed 105~g while the controller-mounted device weighed 41~g or 165~g with the controller.

\subsection{Shared Spring Mechanism}
\label{subs:spring_mechanism}

The spring mechanism provided impulse feedback in both the haptic glove and controller-mounted device.
It was designed to deliver a brief, localized impulse to a contact surface while allowing the solenoid actuation hardware to separate from the hand or controller body. Its design was motivated by prior work showing that brief force or impact cues can enhance contact events in VR, including work on impulse feedback for virtual ball games~\cite{yoshimuraSkinPressuretypeGrasping2022} and solenoid-based tactile stimulation for simulated physical impacts~\cite{lopesImpactoSimulatingPhysical2015}. 

The mechanism consisted of a spring fixed at one end to a rigid support (bridge or casing), with a pan-head screw attached to its free end (Figure~\ref{fig:haptic-modules}). The screw was connected to a fishing line routed through a short, flexible pneumatic tube and into a 1~m stiff Bowden cable to maintain alignment and tension. The other end of the cable was anchored to a remote enclosure, where the fishing line was tied to the plunger of a fixed Heschen push-pull solenoid.

Although both devices used the same cable-driven mechanism, the spring's actuation differed by device. In the haptic glove, powering the solenoid pulled the fishing line (Figure~\ref{fig:haptic-modules}, bottom), bending the spring toward the pneumatic tube opening (Figure~\ref{fig:haptic-modules}, left and middle). This quick motion pressed the rounded spring tip against the target surface (e.g., fingertip), producing a brief, localized impulse sensation. When powered off, line tension was released and the spring returned to its neutral position. In the controller-mounted device, the mechanism operated in the opposite direction. The spring was first pulled away from the controller shell, and when the solenoid was powered off, the spring snapped back and struck the shell (Figure~\ref{fig:haptic-modules}, right), generating an impulse felt by the user through the controller.

\subsection{Haptic Glove Device}
\label{subs:haptic_glove}

The haptic glove was organized around a fingerless glove that served as the palm module and base structure for the hand-worn device. Two separate finger-sleeve modules were worn on the index finger and thumb, and a stretchable white outer glove was worn over all modules to cover and secure the components. This outer glove created a uniform hand silhouette, which improved the headset's hand detection despite the added haptic modules, while preserving finger flexibility with minimal restraint from the fabric. The modular layout also made the glove adjustable. The elastic wrist opening secured the palm module, the finger sleeves used Velcro straps at the base and fingertip, and the spacing between the palm and finger modules could be adjusted to better fit each participant's hand.

\subsubsection{Palm Module}
The palm module used a fingerless glove with a plastic casing affixed at the centre of the palm and oriented parallel to the fingers. The casing housed the spring mechanism, with a coin-type vibration motor beside it (Figure~\ref{fig:haptic-modules}, middle). When grasping, the module delivered impulse or vibrotactile feedback through the casing to the palm and fingers.

\subsubsection{Finger Sleeve Module}
\label{subs:finger_sleeve}

The index finger and thumb used the same finger-sleeve module design. This module ran from the fingertip to the middle knuckle, with the spring mechanism positioned over the finger pad. From the skin outward, the stack consisted of the glove's fabric, a coin-type vibration motor, and the spring mechanism (Figure~\ref{fig:haptic-modules}, left). The module delivered either impulse or vibrotactile feedback to the finger pad when activated.

\subsection{Controller-Mounted Haptic Device}
The controller-mounted module was attached to a Velcro strap wrapped around the controller body without obstructing the {\small\textsc{TRIGGER}} button. The stack consisted of the controller shell, Velcro strap, an inline spring mechanism with two noise-dampening bumpers, and the pneumatic tube (Figure~\ref{fig:haptic-modules}, right). A coin-type vibration motor was secured to the shell adjacent to the spring mechanism using adhesive backing. The module delivered impulse or vibrotactile feedback through the controller shell to the hand.

\subsection{Software Communication}
All haptic modules were driven by a Seeed Studio XIAO ESP32C3 mini development board\footnote{%
Available from \url{https://wiki.seeedstudio.com/XIAO_ESP32C3_Getting_Started/}} programmed using the Arduino framework. Communication between the headset and microcontroller was handled over the same local Wi-Fi network, without relying on external internet access. The ESP32C3 listened for condition-specific trigger messages sent by the Unity software and, upon receiving them, parsed each message to activate the corresponding haptic module. Each module operated at a single haptic feedback intensity, providing binary (on/off) control.

\section{Methodology}

\subsection{Participants}
We recruited 24 participants from the university community through posters and online announcements. Participants were 21 to 47 years old (\textit{M} = 25.5 years, \textit{SD}= 5.2 years). The sample comprised 11 self-identified women (45.8\%), 11 self-identified men (45.8\%), and 2 self-identified non-binary participants (8.3\%). Vision was evenly split between normal and corrected-to-normal (12 each, 50\%). Most participants were right-handed (19, 79.2\%), with 4 left-handed (16.7\%) and 1 ambidextrous (4.2\%). The ambidextrous participant used their right hand.
For VR experience, 5 participants (20.8\%) had never used VR, 15 (62.5\%) had used it once or a few times, and 4 (16.7\%) reported more extensive VR experience. For hand tracking experience, 15 participants (62.5\%) reported no prior experience and 9 (37.5\%) had used it once or a few times. Video gaming experience varied across participants: 3 participants (12.5\%) reported never playing video games, 8 (33.3\%) played less than once every few months, 4 (16.7\%) played monthly, and 9 (37.5\%) played weekly.

\subsection{Apparatus}

\subsubsection{Hardware}

Our study incorporated the custom haptic devices described in Section \ref{sec:hapticdevices}. Participants wore the haptic glove on their dominant hand throughout the experiment, maintaining consistency and reducing session time, while the controller-mounted device remained attached to the dominant-hand controller, which the researcher provided or collected as needed.

We used the Meta Quest 3 head-mounted display. Background music was played through the headset speakers throughout the experiment to reduce participants' exposure to external sounds. Depending on the current input, participants used either the headset's integrated hand tracking or a standard handheld controller to perform the three interactions: grasp, pinch, and tap. Each used either a gesture or a closely matched button press or tap with the controller. With hand tracking, grasp required closing the hand into a fist, and haptic feedback was issued through all glove modules. To perform a pinch, participants brought together the tips of the index finger and thumb and received haptic feedback through both finger-sleeve modules. Tap required a pointing motion with the index finger, and haptic feedback was issued only to the glove's index finger sleeve module. 

In the controller conditions, both pinch and grasp were mapped to the {\small\textsc{TRIGGER}} button. We chose this mapping for several reasons: 1) it more closely resembled the corresponding hand gestures than other button options; 2) it accommodated the haptic module placement; 3) it aligned with the Meta Quest system's primary select action (e.g., menu navigation); and 4) it leveraged participant familiarity with the trigger for selection input. Tap required the virtual controller to intersect the target and did not require a button press, reflecting a common collision-based approach to tapping or poking interactions in VR. All controller-based interactions received haptic feedback through the controller-mounted module.
Because these interactions reflect specific task implementations rather than general grasp, pinch, or tap gestures, we refer to them as grasp-like, pinch-like, and tap-like, and, for clarity, henceforth simply as grasp, pinch, and tap.

\subsubsection{Software}
\begin{figure}[t]
  \centering
  \vspace{0.5em}

  \begin{subfigure}[]{1.0\columnwidth}
    \centering
    \includegraphics[width=\linewidth]{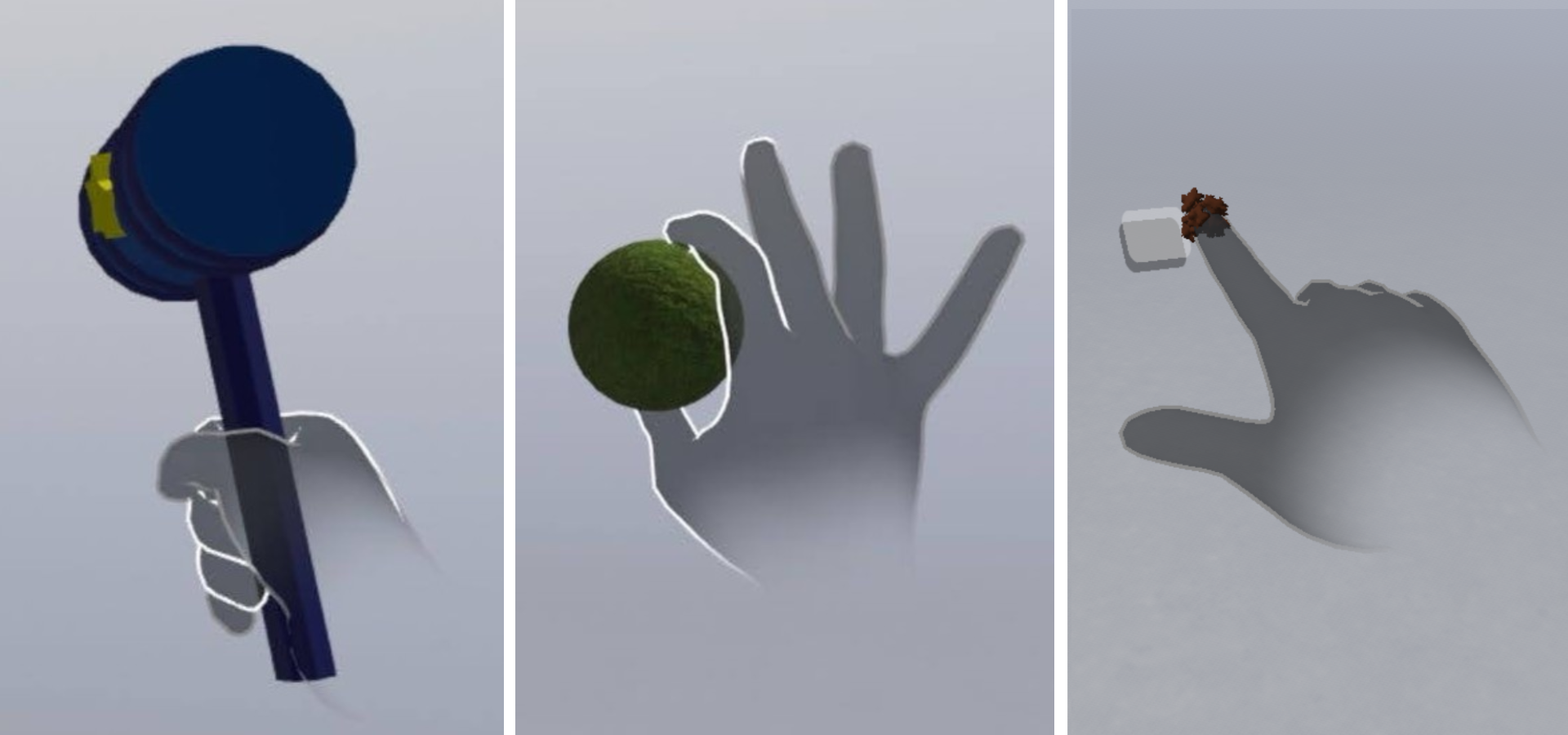}
    \label{subfig:hand_interactable}
  \end{subfigure}

  \vspace{-0.8em}

  \begin{subfigure}[]{1.0\columnwidth}
    \centering
    \includegraphics[width=\linewidth]{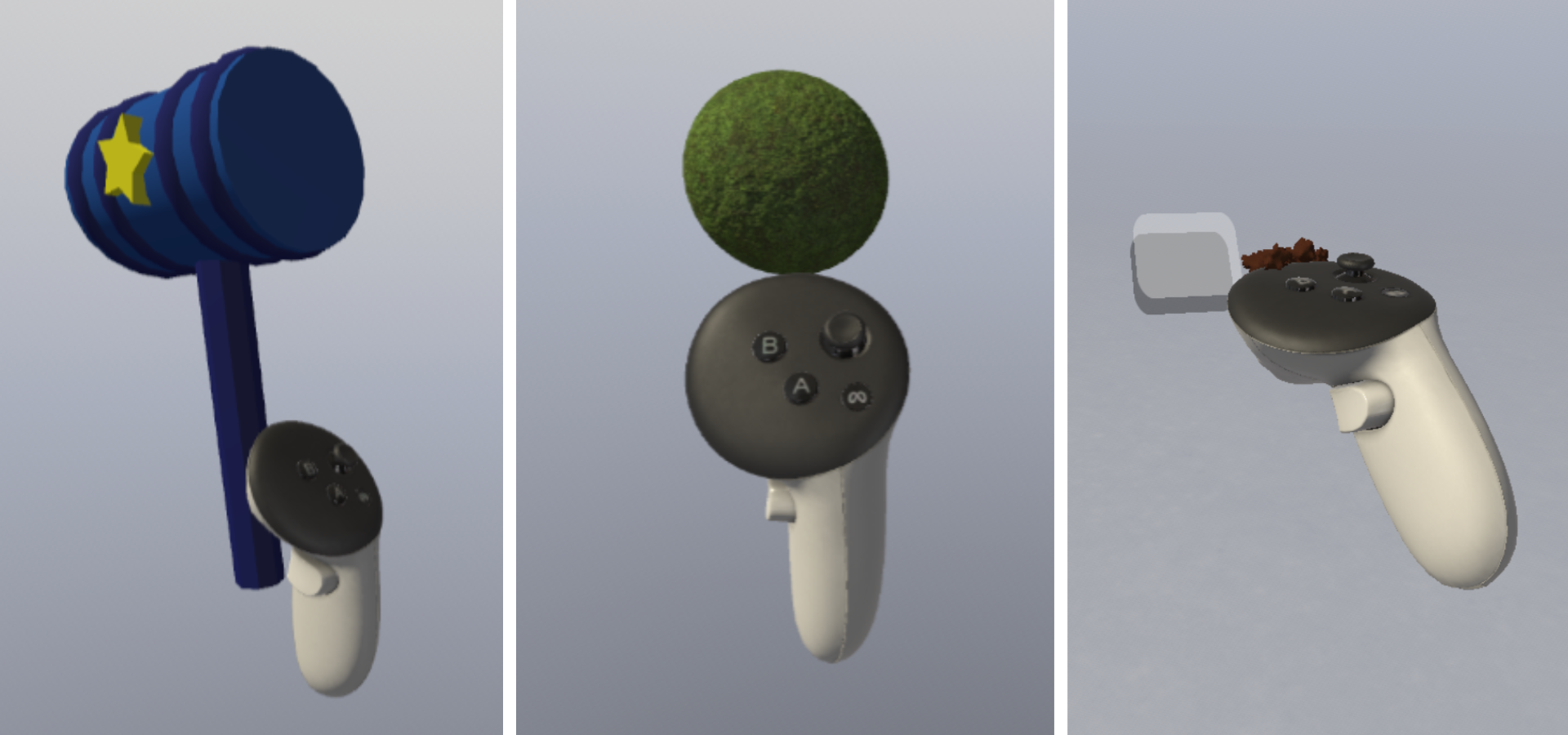}
    \label{subfig:controller-interactable}
  \end{subfigure}

  \vspace{-1.2em}
  \caption{Interactable objects. The first row shows the objects for the \textit{hand tracking} conditions: a hammer (grasp), a seed (pinch), and a button (tap). The second row shows the corresponding objects in the \textit{controller} conditions.}
  \label{fig:interactable-objects}
\end{figure}

We used Unity 2022.3.1f1 and the Meta XR All-in-One SDK to develop the study's VR software. The virtual environment (VE) was adapted from Ocampo et al.'s earlier study~\cite{ocampoComparingHandController2025}, retaining an open, low-detail layout to reduce distraction.
The evaluation area consisted of a static planting bed positioned directly in front of the participant and slightly tilted toward them to support a comfortable interaction posture.
It displayed instructions and provided the shared interaction surface for the study tasks. 
The virtual camera was positioned at the centre of the room, 3~m above the floor, to prevent height misalignment between the seated participant and the evaluation space upon entering the VE. Participants saw a hand avatar when using hand tracking and a controller avatar when using controllers (Figure~\ref{fig:interactable-objects}). Providing a visible input representation can improve target-selection performance~\cite{demarbre2025effects}, while matching that representation to the input may further improve performance and user experience~\cite{ocampoComparingHandController2025}. 
The software also controlled trial progression and recorded all dependent variables in real time to a Google Sheets spreadsheet via an internet connection.

\subsubsection{Task Apparatus}
All tasks used the same target-acquisition layout to support cross-condition comparison. This included six circularly arranged target rings, a central starting point, and an object-spawn location above the starting point. Each ring was 0.11~m $\times$ 0.07~m $\times$ 0.11~m, with neighbouring rings separated by 25~cm. Each task used a task-specific object matched to the intended interaction: a hammer for grasp, a seed for pinch, and a button for tap.

Haptic feedback was delivered by the appropriate device module in response to software events triggered by participants' interactions with task objects, such as striking a target with the virtual hammer. The specific events that triggered haptic feedback are described in the task descriptions in Section~\ref{subs:procedure_ch}.

\begin{figure*}[!t]
  \centering
  \begin{subfigure}[]{0.323\textwidth}
    \centering
    \includegraphics[width=\linewidth]{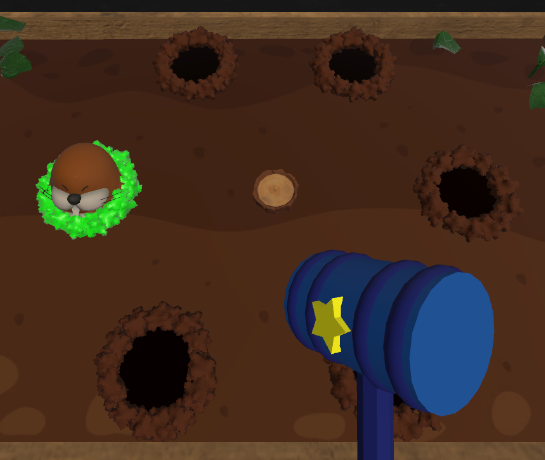}
    \caption{Grasp task}
    \label{fig:grasp-task}
  \end{subfigure}
  \hfill
  \begin{subfigure}[]{0.323\textwidth}
    \centering
    \includegraphics[width=\linewidth]{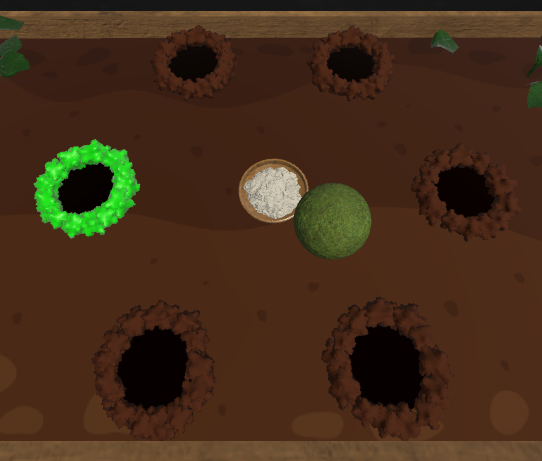}
    \caption{Pinch task}
    \label{fig:pinch-task}
  \end{subfigure}
  \hfill
  \begin{subfigure}[]{0.334\textwidth}
    \centering
    \includegraphics[width=\linewidth]{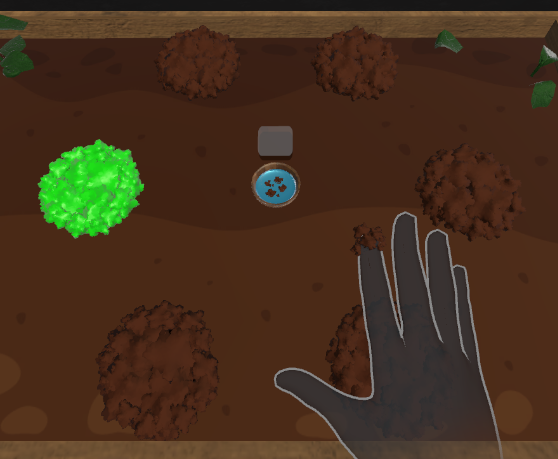}
    \caption{Tap task}
    \label{fig:tap-task}
  \end{subfigure}
    \vspace{-1.2em}
  \caption{Tasks used in the study: a) grasp, b) pinch, and c) tap.}
  \label{fig:tasks}
\end{figure*}

\subsection{Procedure}
\label{subs:procedure_ch}
The study protocol was approved by our institutional research ethics board. Participants first provided informed consent and then completed a demographic questionnaire. Next, the researcher demonstrated the three tasks, two input modalities, and three haptic conditions before the experiment began. 
During the experimental trials, participants performed the gesture-specific tasks (grasp, pinch, or tap). Across gestures, target order always followed a sequence inspired by the ISO 9241-411 reciprocal selection task \cite{ISOTS92414112012}, beginning with the top-right ring and proceeding to the next target on the opposite side of the target arrangement. Regardless of task, each trial consisted of three phases: 1) interaction with an object at the object-spawn-point; 2) interaction with the centre of the ring arrangement (hereafter referred to as the centre-point); and 3) interaction with the target ring. These phases are described in the task descriptions below. After each trial, the system reset the task for the next trial. 

With each new condition, participants completed 3 non-recorded practice trials, followed by 24 recorded trials.
After each condition, participants completed an in-VR questionnaire adapted from the NASA-TLX, with its design informed by prior work on in-VR questionnaire design~\cite{kuntzerTellingUsWhat2026}. It included six workload items rated on a 5-point Likert scale (1 = low, 5 = high): physical demand, mental demand, temporal demand, perceived success, effort, and frustration. Higher ratings indicated greater workload except for perceived success, for which higher ratings were more favourable. The questionnaire also included three preference items in which participants rated how much they liked or disliked the input, gesture, and haptic feedback in that condition on a 5-point Likert scale (1 = strongly dislike, 5 = strongly like). Because these ratings were collected after each condition, participants rated the same input, gesture, or haptic feedback multiple times across different combinations of the other experimental factors.
Participants were free to take breaks between conditions. At the end of the experiment, participants completed a post-study questionnaire about their overall experience and preferences before receiving CAD~\$15 in compensation.
The tasks included:

\par\smallskip
\noindent\textbf{{Grasp Task (\textit{Whack-a-Mole}): }}
Participants grabbed a virtual hammer to hit moles that popped up from target rings. At the start of each trial, a hammer spawned at the object-spawn-point. Participants grasped its handle using a grasp gesture with the current input. %
Participants then struck the tree stump at the centre-point, triggering haptic feedback. Upon hitting the tree stump, the next target ring activated and turned green, and a 3D model of a mole appeared\footnote{The 3D mole model in the target ring was retrieved from \href{https://skfb.ly/oKNQp}{https://skfb.ly/oKNQp}.}. Participants then moved the hammer to the target ring and struck the mole's head to complete the trial. Striking the mole with the hammer triggered a second haptic feedback event. %
See Figure~\ref{fig:grasp-task}.

\par\smallskip

\noindent\textbf{{Pinch Task (\textit{Plug-a-Hole}): }}
Participants pinched a round seed to plant it in target rings. At the start of each trial, the seed spawned at the object-spawn-point. Participants acquired the seed using a pinch gesture with the assigned input. They then dipped it into the bowl of fertilizer at the centre-point, which activated the target ring and turned it green. Touching the seed to the fertilizer triggered haptic feedback. Participants then moved the seed to the target ring and inserted it (plugging the hole) to complete the trial. Contact between the seed and target ring triggered haptic feedback a second time. %
See Figure~\ref{fig:pinch-task}.
\par\smallskip

\noindent\textbf{{Tap Task (\textit{Poke-a-Hole}): }}
Participants tapped a button to poke planting holes in target rings. At the start of each trial, a button spawned at the object-spawn-point. Participants pressed the button with a tap using the assigned input. After a successful press, the button disappeared, enabling participants to interact with the water basin at the centre-point. Participants then tapped the water basin, turning the target ring green and triggering haptic feedback. Next, participants tapped the target ring to complete the trial. Contact with the target ring triggered haptic feedback a second time. %
See Figure~\ref{fig:tap-task}.

\subsection{Design}
The experiment employed a $2\times3\times3$ within-subjects design. The independent variables and their levels were:

\begin{itemize}
  \item \textbf{Input}: Controller, Hand Tracking 
  \item \textbf{Haptic Feedback}: Vibration, Impulse, None 
  \item \textbf{Gesture Type}: Grasp, Pinch, Tap
\end{itemize}

The order of the 18 combinations of these factors was counterbalanced according to a balanced Latin square. Participants completed 24 recorded trials for each condition for a total of 432 trials (2 × 3 × 3 = 18 conditions $\times$ 24 trials), consisting of 144 trials per gesture (excluding practice trials). The experiment took an average of around 55 minutes to complete.

Our dependent variables included movement time and contact deviation. Movement time, measured in seconds, was the time from the initial interaction with the centre-point to successful interaction with the target. Contact deviation, measured in cm, was the distance from the target ring’s centre to the task object's centre for grasp (hammer) and pinch (seed), or to the participant's finger pad or controller tip for tap. The collected data are available on the \href{https://osf.io/r385v/overview?view_only=6da6a533772640c7bae045381925fd6a}{Open Science Framework}\footnote{OSF \url{https://osf.io/r385v/overview?view_only=6da6a533772640c7bae045381925fd6a}}.

\section{Results}

\begin{figure*}[htbp!]
  \centering
  \includegraphics[width=\textwidth]{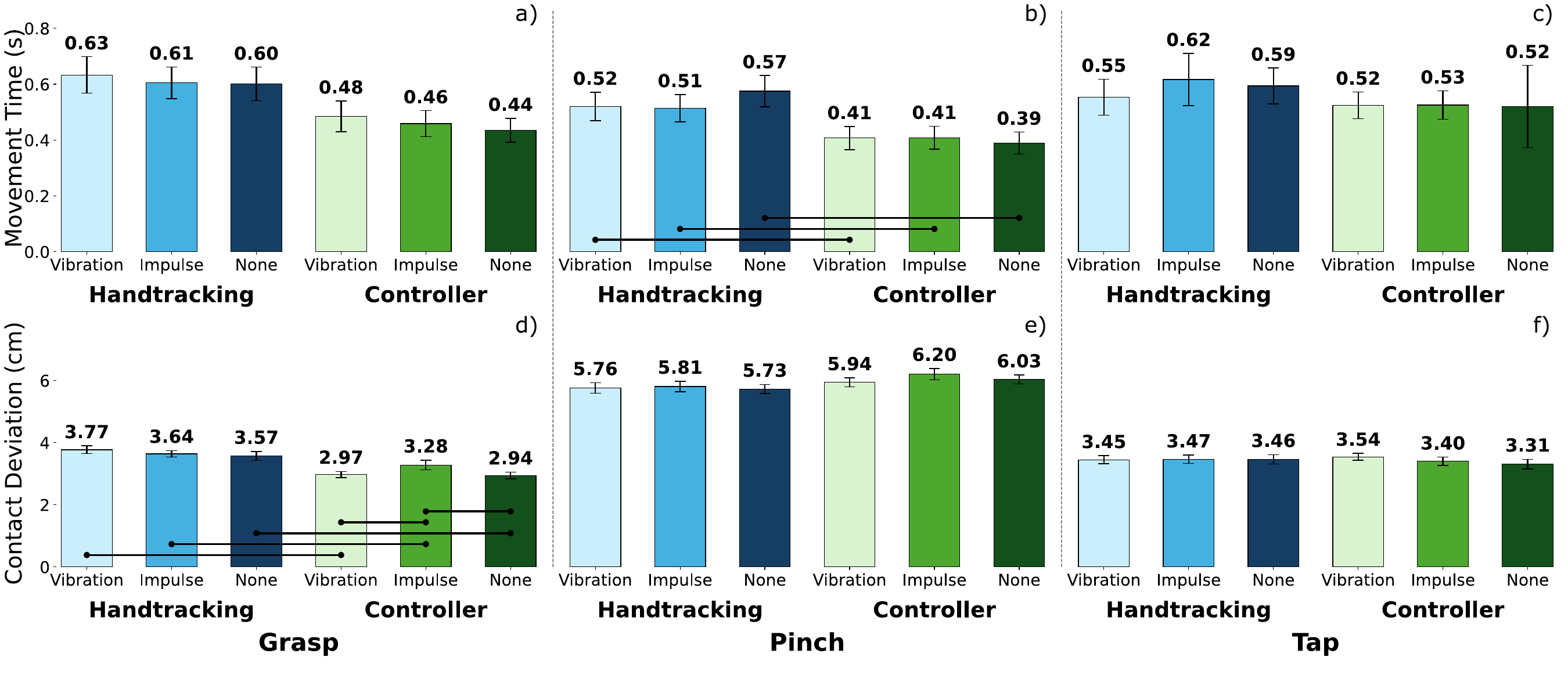}
  \caption{Mean movement time in seconds (first row) and mean contact deviation in centimetres (second row) across input and haptic feedback. \textbf{Left: }Grasp (Whack-a-Mole), \textbf{Centre: }Pinch (Plug-a-Hole), \textbf{Right: } Tap (Poke-a-Hole). Error bars show ±1 SE.}
  \label{fig:result-all-t}
\end{figure*}

We analyzed objective performance separately for each task using a $2\times3$ within-subjects repeated-measures ANOVA with factors input and haptic feedback. These analyses were conducted for movement time and contact deviation in the grasp (\textit{Whack-a-Mole}), pinch (\textit{Plug-a-Hole}), and tap (\textit{Poke-a-Hole}) tasks. For subjective measures, we used Aligned Rank Transform (ART) ANOVAs~\cite{wobbrockAlignedRankTransform2011}. Each NASA-TLX item was analyzed separately for each gesture using a $2\times3$ ART ANOVA with factors input and haptic feedback. The three condition-level evaluative ratings were analyzed using a $3\times3\times2$ ART ANOVA with factors gesture, haptic feedback, and input. Post-hoc comparisons used Bonferroni correction for the objective and NASA-TLX analyses, and Holm correction for the evaluative ratings. In Figure~\ref{fig:result-all-t}, error bars show $\pm 1$ SE, and black horizontal lines between bars indicate significant pairwise comparisons after adjustment ($p < .05$).

\subsection{Grasp Task (Whack-a-Mole)}

\textbf{Movement Time: }There was a significant main effect of input on movement time ($F_{1,23}=25.27$, $p<.001$, $\eta_p^2=.52$), favouring controllers (mean 0.52~s, \textit{SD} 0.22~s) over hand tracking (mean 0.68~s, \textit{SD} 0.28~s), as shown in Figure~\ref{fig:result-all-t}a. Neither the main effect of haptic feedback ($F_{1.41,32.52}=2.89$,$p=.086$,$\eta_p^2=.112$) nor the input $\times$ haptic feedback interaction effect ($F_{1.35,31.02}=0.27$,$p=.681$,$\eta_p^2=.011$) were significant, suggesting that movement time differed by input, but not reliably by haptic feedback or its interaction with input.

\textbf{Contact deviation: } There was a significant main effect of input on contact deviation ($F_{1,23}=25.59$, $p<.001$, $\eta_p^2=.53$). Controller conditions produced lower contact deviation (mean 3.1~cm, \textit{SD} 0.62~cm) than hand tracking (mean 3.7~cm, \textit{SD} 0.61~cm), indicating contacts were closer to the target centre with controllers.
Haptic feedback was also significant ($F_{2,46}=3.83$, $p=.029$, $\eta_p^2=.14$). The none condition produced lower contact deviation (mean = 3.31~cm, \textit{SD} = 0.53~cm) than impulse (mean 3.52~cm, \textit{SD} 0.54~cm).
The input $\times$ haptic feedback interaction was also significant ($F_{2,46}=5.65$, $p=.006$, $\eta_p^2=.20$). Post-hoc tests revealed that hand tracking produced higher contact deviation than controllers within each haptic feedback condition, consistent with the main effect of input. 
In the controller conditions, impulse produced significantly higher contact deviation (mean 3.4~cm, \textit{SD} 0.76~cm) than vibration (mean = 3.0~cm, \textit{SD} = 0.48~cm) or no haptic feedback (mean = 3.0~cm, \textit{SD} = 0.54~cm; Figure~\ref{fig:result-all-t}d). 
Despite these significant differences, the observed effects were small in absolute terms (i.e., few millimetres).

 \subsection{Pinch Task (Plug-a-Hole)}

\textbf{Movement Time.} For pinch (Figure~\ref{fig:result-all-t}b), there was a significant main effect of input on movement time ($F_{1,23}=39.84$,$p<.001$,$\eta_p^2=.63$), with faster movements using controllers (mean of 0.45~s, \textit{SD} = 0.19~s) than hand tracking (mean of 0.60~s, \textit{SD} = 0.24~s). The main effect of haptic feedback was not significant ($F_{1.55,35.70}=1.17$,$p=.311$,$\eta_p^2=.048$). However, the interaction effect input $\times$ haptic feedback was significant ($F_{1.39,31.90}=4.28$,$p=.035$,$\eta_p^2=.16$). Controllers were faster than hand tracking across all three haptic feedback conditions, consistent with the main effect. The gap was largest with no haptic feedback (i.e., hand tracking $\times$ none had a mean of 0.64~s, \textit{SD} = 0.27~s vs. controller $\times$ none had a mean of 0.44~s) and smaller with either vibration or impulse (hand tracking $\times$ vibration had a mean of 0.59~s, \textit{SD} = 0.25~s; hand tracking $\times$ impulse had a mean of 0.58~s, \textit{SD} = 0.24~s; controller conditions remained 0.44-0.46~s).

\textbf{Contact Deviation.} There was a significant main effect of input on contact deviation ($F_{1,23}=5.60$,$p=.027$,$\eta_p^2=.20$). In contrast to grasp, hand tracking offered slightly lower (i.e., more accurate) contact deviation (mean of 5.8~cm, \textit{SD} = 0.74~cm) than controllers (mean of 6.1~cm, \textit{SD} = 0.67~cm), as seen in Figure~\ref{fig:result-all-t}e. Neither haptic feedback ($F_{2,46}=2.17$,$p=.126$,$\eta_p^2=.086$) nor the interaction ($F_{2,46}=1.39$,$p=.26$,$\eta_p^2=.057$) yielded significant effects, indicating that precision with pinch was largely driven by the input.

\subsection{Tap Task (Poke-a-Hole)}

\textbf{Movement Time.} For the tap task, (Figure~\ref{fig:result-all-t}c), we found no significant effects for input ($F_{1,23}=1.05$,$p=.316$,$\eta_p^2=.044$), haptic feedback ($F_{1.25,28.84}=0.75$,$p=.422$,$\eta_p^2=.032$), nor their interaction ($F_{1.42,32.66}=.80$,$p=.421$,$\eta_p^2=.033$).

\textbf{Contact Deviation.} Much like movement time, there were no significant effects for input ($F_{1,23}=.12$,$p=.729$,$\eta_p^2=.005$), haptic feedback ($F_{2,46}=.63$,$p=.535$,$\eta_p^2=.027$), nor their interaction ($F_{2,46}=1.76$,$p=.183$,$\eta_p^2=.071$) for contact deviation (Figure~\ref{fig:result-all-t}f).

\subsection{In-study NASA-TLX}
\begin{figure*}[!thp]
  \centering
  \includegraphics[width=\linewidth]{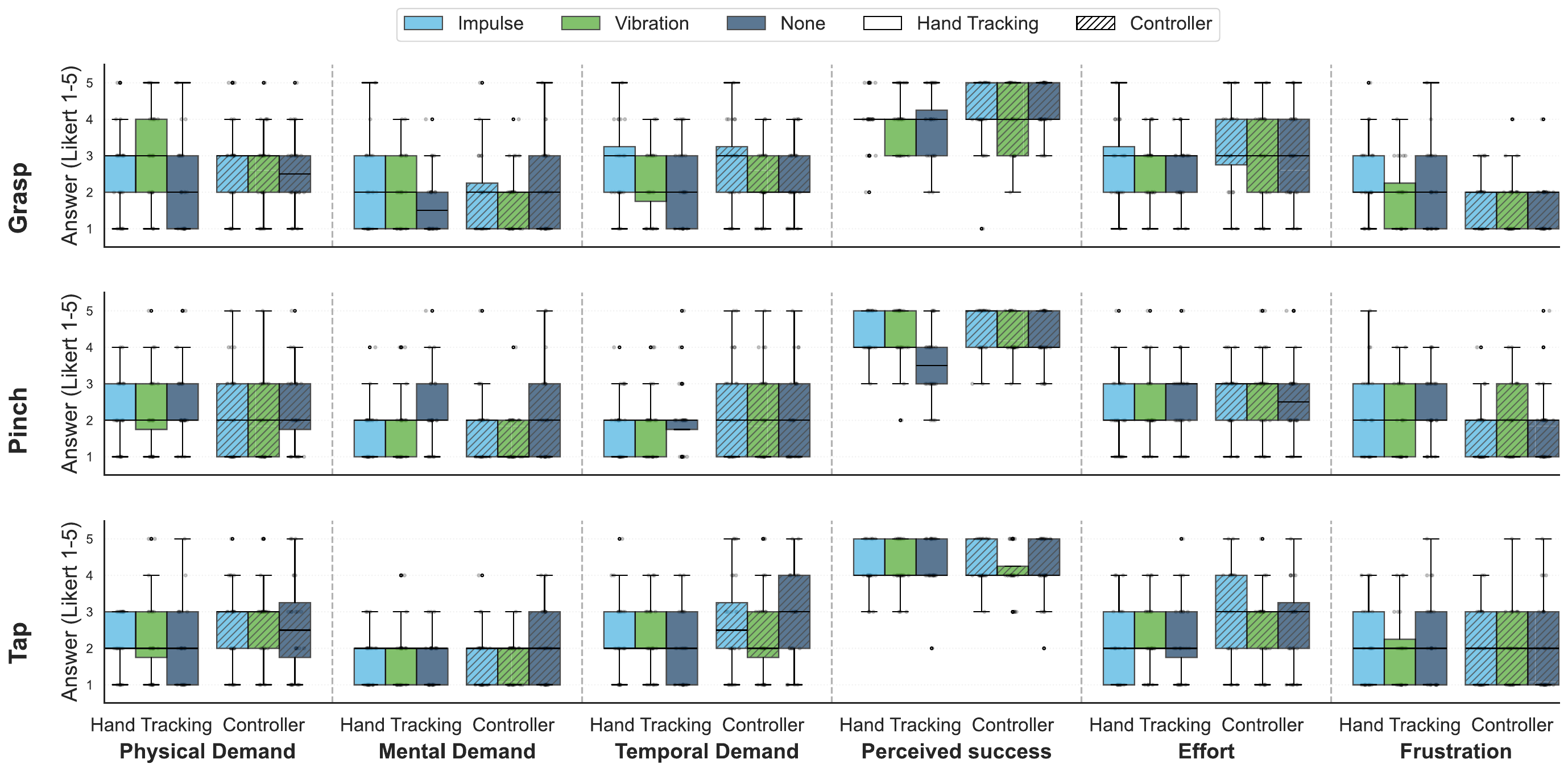}
  \caption{NASA-TLX workload ratings for each gesture by input and haptic feedback condition.}
  \label{fig:result-grasp-a}
\end{figure*}

Figure~\ref{fig:result-grasp-a} shows the distribution of NASA-TLX workload ratings across gestures, input, and haptic feedback conditions.

\textbf{Grasp Task (Whack-a-Mole).} Workload ratings showed several significant effects. Physical demand varied by haptic feedback ($F_{2,115}=4.18$, $p=.018$), with vibration rated higher than no feedback (mean scores of 2.83 vs. 2.40). Mental demand showed a main effect of input ($F_{1,115}=4.18$, $p=.043$), with controllers rated slightly higher than hand tracking (2.04 vs. 1.99), and an input $\times$ haptic feedback interaction ($F_{2,115}=3.37$, $p=.038$). Within vibration, hand tracking was rated higher than controllers (2.13 vs. 1.88), while within controllers, no feedback was rated higher than impulse and vibration (2.17 vs. 2.08 and 1.88). Temporal demand varied by haptic feedback ($F_{2,115}=8.76$, $p<.001$), with impulse rated higher than no feedback and vibration (2.79 vs. 2.33 and 2.25). Perceived success showed a main effect of input ($F_{1,115}=5.98$, $p=.016$), with controllers rated higher than hand tracking (4.08 vs. 3.89). Frustration showed main effects of input ($F_{1,115}=40.12$, $p<.001$) and haptic feedback ($F_{2,115}=4.89$, $p=.009$): controllers were less frustrating than hand tracking (1.75 vs. 2.19), and impulse was more frustrating than no feedback and vibration (2.15 vs. 1.92 and 1.85).

\textbf{Pinch Task (Plug-a-Hole).} Workload ratings were affected by both input and haptic feedback. Mental demand varied by haptic feedback ($F_{2,115}=6.04$, $p=.003$), with no feedback rated higher than impulse and vibration (2.19 vs. 1.77 and 1.71). Temporal demand showed a main effect of input ($F_{1,115}=13.86$, $p<.001$), with controllers rated higher than hand tracking (2.35 vs. 1.96). Perceived success showed main effects of input ($F_{1,115}=18.87$, $p<.001$) and haptic feedback ($F_{2,115}=14.58$, $p<.001$), and an input $\times$ haptic feedback interaction ($F_{2,115}=9.22$, $p<.001$). Controllers were rated higher than hand tracking overall (4.31 vs. 4.08), and no feedback was rated lower than impulse and vibration (3.85 vs. 4.42 and 4.31). Within no feedback, controllers were rated higher than hand tracking (4.17 vs. 3.54), while within hand tracking, vibration was rated higher than no feedback (4.33 vs. 3.54). Effort showed a main effect of input ($F_{1,115}=11.53$, $p<.001$), with controllers rated higher than hand tracking (2.71 vs. 2.44). Frustration also showed a main effect of input ($F_{1,115}=6.44$, $p=.012$), with hand tracking rated as more frustrating than controllers (2.22 vs. 1.90), and an input $\times$ haptic feedback interaction ($F_{2,115}=3.81$, $p=.025$).

\textbf{Tap Task (Poke-a-Hole).} Workload ratings were mainly affected by input. Controllers were rated higher than hand tracking for physical demand ($F_{1,115}=8.29$, $p=.005$; 2.67 vs. 2.22), mental demand ($F_{1,115}=8.44$, $p=.004$; 1.82 vs. 1.74), temporal demand ($F_{1,115}=9.45$, $p=.003$; 2.65 vs. 2.24), and effort ($F_{1,115}=15.72$, $p<.001$; 2.75 vs. 2.26). Mental demand also showed an input $\times$ haptic feedback interaction ($F_{2,115}=6.99$, $p=.001$): within hand tracking, vibration was rated higher than impulse and no feedback (1.83 vs. 1.67 and 1.71). Temporal demand showed the same interaction pattern ($F_{2,115}=5.03$, $p=.008$), with vibration rated higher than no feedback within hand tracking (2.33 vs. 2.00).%

\subsection{Condition-level evaluative ratings}

\begin{figure}[htpb!]
  \centering
  \includegraphics[width=\linewidth]{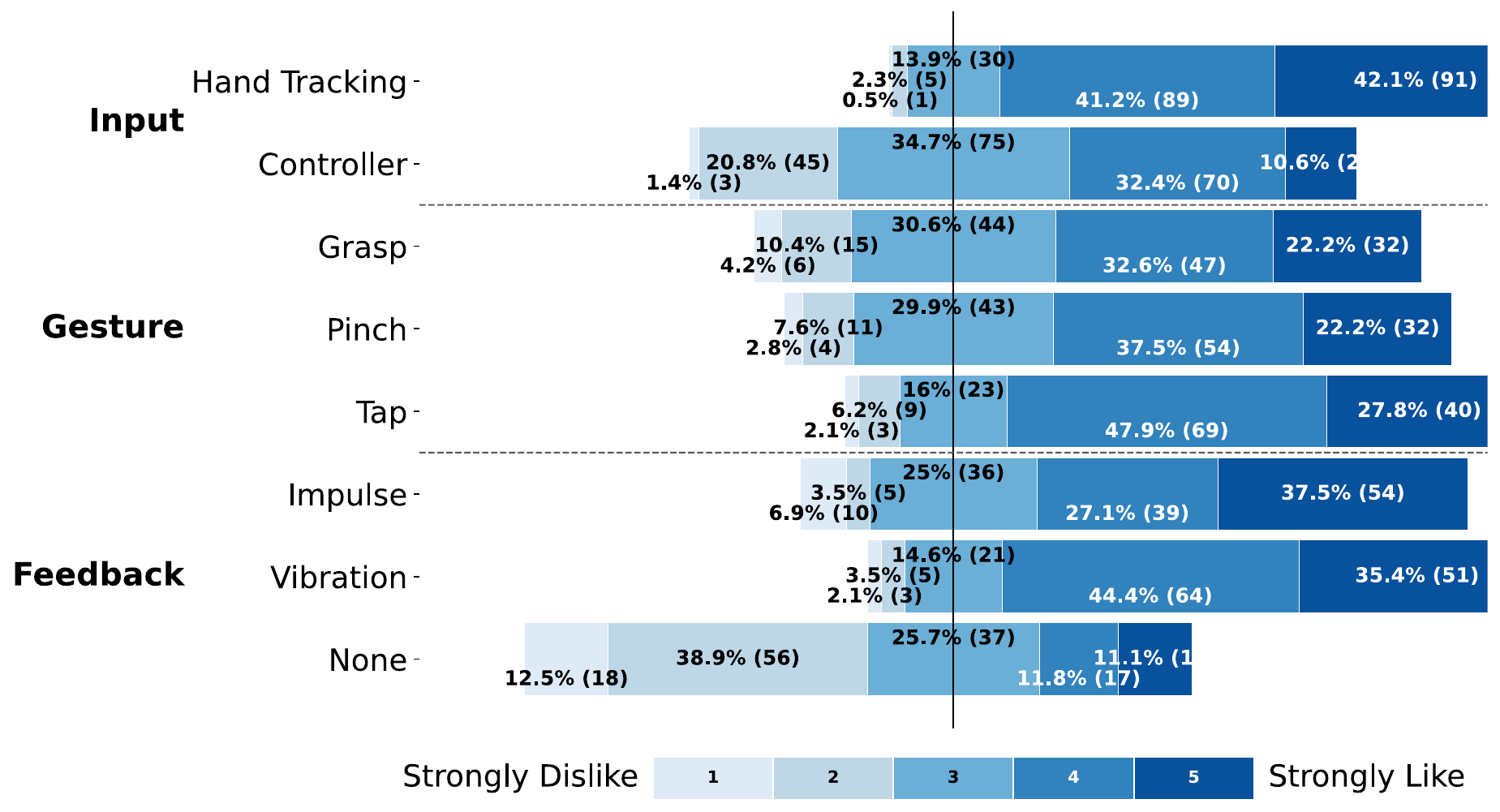}
  \caption{Distribution of evaluative ratings for the input, gesture, and haptic feedback experienced in each condition.}
  \label{fig:preferences}
\end{figure}

Figure~\ref{fig:preferences} shows the distribution of evaluative ratings collected after each condition. The strongest pattern was the effect of input. Hand tracking received higher input ratings than controllers ($F_{1,391}=208.81$, $p<.001$; 83.33\% vs. 43.06\% positive ratings). This trend also appeared in the gesture and feedback ratings. Gesture ratings were affected by gesture ($F_{2,391}=11.36$, $p<.001$) and input ($F_{1,391}=65.06$, $p<.001$): tap was rated higher than grasp and pinch (75.70\%, 54.86\%, and 59.72\% positive ratings), and gestures were rated higher with hand tracking than controllers (73.61\% vs. 53.24\%). The gesture $\times$ input interaction was also significant ($F_{2,391}=4.02$, $p=.019$). Feedback ratings were affected by feedback ($F_{2,391}=102.78$, $p<.001$) and input ($F_{1,391}=9.60$, $p=.002$): no feedback was rated lower than impulse and vibration (22.92\%, 64.58\%, and 79.86\% positive ratings), and the feedback $\times$ input interaction was significant ($F_{2,391}=11.03$, $p<.001$). Overall, hand tracking was evaluated more positively, and this input trend also shaped how participants rated gestures and haptic feedback.

\subsection{Post-study preference summary}

Post-study preference answers followed the same general trends. Hand tracking was selected as the preferred input by 19 of 24 participants (79\%), mainly due to comfort (18/19, 95\%), directly using one's own hand (17/19, 89\%), and ease of use (16/19, 84\%). As P3 explained, hand tracking meant they did not have to "think which button to press" and could instead use their "actual hand gestures." Controllers were preferred by 5 participants (21\%), mainly due to responsiveness and perceived task performance (4/5, 80\% each).

Participants also favoured active haptic feedback. Vibration was ranked as the most realistic feedback type by 11 participants (46\%) and impulse by 10 (42\%), whereas no feedback was selected by only 3 participants (13\%). Haptic feedback was reported to enhance enjoyment (20/24, 83\%), realism (17/24, 71\%), performance (16/24, 67\%), task focus (15/24, 63\%), and confidence (15/24, 63\%). Participants described vibration as useful for confirming contact, while impulse conveyed a stronger hit-like sensation. For example, P19 noted that vibration made it easier to know they ``did it right,'' whereas P11 described impulse as providing ``the natural feeling associated with hitting something.'' However, three participants described impulse feedback as strange, distracting, or uncomfortable.

Tap was most preferred and required least effort for both controllers (14/24 preferred; 16/24 least effort) and hand tracking (13/24 preferred; 19/24 least effort). Comments reflected this pattern: P11 called tap "the most intuitive gesture," while P19 said it "required less effort with hand tracking." Participants most often chose tap with vibration for both controllers (11/24, 46\%) and hand tracking (8/24, 33\%) as their preferred gesture-haptic pairings.

\section{Discussion}

\subsection{User Performance}

Across gestures, performance was influenced more strongly by input than haptic feedback. 
Grasp and pinch were faster with controllers than hand tracking. The controller input modality was also more accurate, with lower contact deviation with grasp. This pattern is consistent with prior work that found that controller-based input offers faster and more precise task execution than free-hand interaction \cite{buckinghamHandTrackingImmersive2021a, kangasTradeOffTaskAccuracy2022, luongControllersBareHands2023}. However, this result was not consistent across all gestures or performance measures. For pinch, hand tracking yielded a small but significant reduction in contact deviation despite controllers remaining faster overall. This suggests hand tracking may offer better accuracy for interactions relying on direct fingertip control. In contrast, grasp required participants to control a distal contact point (i.e., the virtual hammer), which likely increased alignment demands and limited the contact-deviation advantage observed for hand tracking in pinch. Pinch had higher contact deviation than the other gestures across inputs, suggesting additional challenges for precise target contact. However, this effect may also have been influenced by the different objects and target configurations in each task.

The effect of haptic feedback was more limited and gesture-dependent. For movement time, only the input $\times$ haptic feedback interaction was significant, and only for the pinch gesture. However, the follow-up comparisons mainly reflected the same input pattern observed in the main effect, with pairwise differences favouring controllers over hand tracking, rather than revealing clear differences among haptic feedback conditions. No haptic feedback effects were found for movement time in the other gestures.

For contact deviation, haptic feedback had an effect only in the grasp task. The no-feedback condition had lower contact deviation than impulse, and the input $\times$ haptic feedback interaction showed that hand tracking had higher contact deviation than controllers for each haptic feedback condition, consistent with the main effect of input. With controllers, impulse produced higher contact deviation than both vibration and no haptic feedback. These differences were small in absolute terms, 
but they suggest that haptic feedback is not automatically beneficial for precision. A possible explanation is that the impulse cue delivered through the controller-mounted device may have affected the final contact position, an example of the so-called Heisenberg effect which is known to have an impact on selection accuracy \cite{wolfUnderstandingHeisenbergEffect2020}. This effect may have mattered more in the grasp task because participants had to guide a virtual hammer toward a distal target, where small variations in the controller's final position could be reflected at the hammer endpoint. This interpretation remains tentative, but it highlights a potential trade-off between physically expressive feedback and precise endpoint control.

Finally, tap showed little difference across input, haptic feedback, or their interaction. This may indicate that objective tap performance was less sensitive to these manipulations, or that the task was simple enough for participants to perform similarly across input and feedback conditions.

\subsection{User Experience}

Subjective ratings and post-study preferences contrasted with the objective performance results. Despite controllers' superior objective performance, hand tracking was rated more positively and was preferred in the post-study survey. DeMarbre et al.~\cite{demarbre2025effects} report a similar divergence: some participants preferred no controller representation in MR, although the absence of a hand or controller representation yielded poorer selection performance. Hand tracking was mainly associated with comfort, ease of use, and the feeling of directly using one's own hand, whereas the smaller group who preferred controllers emphasized responsiveness and perceived task performance. Workload ratings were mixed (Figure~\ref{fig:result-grasp-a}): controllers were rated as less frustrating and more successful for grasp, but more rushed and effortful with pinch, and more physically demanding, mentally demanding, rushed, and effortful with tap. Thus, faster interaction did not consistently translate to lower perceived workload or better subjective preference, consistent with the trade-off between interaction efficiency and perceived naturalness described in prior work \cite{kangasTradeOffTaskAccuracy2022, luongControllersBareHands2023}.
Input also influenced gesture and haptic feedback ratings, suggesting that participants' experience depended partly on whether they used controllers or hand tracking.

Haptic feedback had a greater impact on the subjective metrics than user performance. Both haptic feedback options were better rated than no feedback. Vibration and impulse were linked to enjoyment, realism, perceived performance, focus, and confidence. This was particularly evident using hand tracking for pinch, where no haptic feedback yielded lower perceived success, especially compared to vibration. These results suggest some benefit of haptic feedback in hand-tracking interactions, although specific workload effects depend on the task and cue. This is consistent with Moon et al. \cite{moonHandTrackingVibrotactile2023, moonEffectsWholeHandInteractions2023}, who found that vibrotactile augmentation improved subjective aspects of VR interaction, including presence and engagement.

Vibration and impulse appeared to support the experience in different ways. Vibration was described as a familiar confirmation cue, while impulse was associated with a more physical hit-like sensation, consistent with prior work \cite{yoshimuraSkinPressuretypeGrasping2022}. However, impulse was not consistently favoured over vibration and, in some cases, increased workload or discomfort. This suggests that more physically expressive feedback may increase realism in some contexts, but may also introduce discomfort or distraction.

Gesture responses further suggest that subjective experience depended on the interaction context. Tap was most positively rated and least effortful in post-study rankings, although NASA-TLX ratings still revealed higher workload for controllers with tap. Pinch and grasp were more variable: pinch favoured hand tracking for some workload items, while controllers had higher perceived success and lower frustration; grasp favoured controllers in both objective performance and some workload measures.

Overall, input was the most influential factor, but its effect differed across objective performance and subjective experience. Controllers supported faster movement in grasp and pinch, and lower contact deviation in grasp, while hand tracking showed lower contact deviation in pinch. Subjectively, the pattern moved in the opposite direction: hand tracking was rated more positively overall, and input also shaped ratings of gestures and haptic feedback. Thus, the main trade-off was between objective efficiency and perceived naturalness, comfort, and workload, depending on the gesture.

Although this study used controlled interaction tasks resembling common VR gestures rather than a specific application, our findings suggest that the input device and haptic feedback choices should depend on a given application's interaction demands rather than be treated as one universal configuration. For time-critical tasks involving acquisition and movement of objects, such as warehouse or training simulations, controllers offer faster movement times when used with grasp and pinch-like operations. For high-precision tasks, such as carefully fitting small parts together, hand tracking offers lower contact deviation when used with pinch, suggesting it is well-suited to accuracy-demanding tasks. Applications requiring prolonged engagement, such as therapeutic or social/gaming experiences, may instead prioritize comfort over speed or accuracy; in this respect, hand tracking was consistently rated more favourably. Haptic feedback, especially vibration, improved subjective experience across these contexts. Given the ease with which it can be incorporated into contemporary VR systems, we argue it may be valuable to include, though its objective performance benefits are task-dependent.
These results can help guide design decisions across various VR applications, without prescribing a one-size-fits-all setup for VR interaction design.

\section{Limitations and Future Work}

Several limitations should be noted. 
First, despite our efforts to improve consistency between haptic feedback, there were still some differences between input modalities. We also note that we did not directly measure transmitted vibration or impulse intensity. 
Thus, haptic feedback was likely not fully equivalent across inputs: the glove delivered feedback directly to the fingertips and palm through fabric, whereas the controller transmitted it through the plastic shell. Differences in material and delivery path may therefore have affected perceived feedback.
Gesture-to-actuator mapping also differed across inputs. Hand-tracking gestures activated different numbers of actuators, whereas controller gestures always used one. Consequently, the reported input $\times$ gesture effects may partly reflect actuator mapping rather than gesture differences alone.
Second, controller gestures only approximated their hand-tracking counterparts, limiting direct performance comparisons across inputs. Neither a trigger press nor a controller collision reproduces the same motion as closing a fist, pinching the fingers together, or pointing with the index finger, which limits how directly gesture performance can be compared across inputs.
Third, haptic feedback used a single binary intensity, so results may not generalize to other strengths. %
The fabric glove may also have affected controller grip, although no participants reported difficulty or dropped the controller. Finally, the impulse mechanism produced an audible metallic sound that was not fully masked by background music, potentially adding an auditory cue to the haptic experience.

Future work should directly measure haptic timing and transmitted feedback intensity across devices, reduce the audible noise of the impulse mechanism (e.g., solenoid padding or noise-cancelling headphones), and evaluate additional feedback types and intensities. Because participants wore the glove during the no-feedback condition, we could not isolate haptic feedback's subjective benefits from the "cost" of wearing the glove itself. Future work should therefore include a glove-free condition to compare bare-handed tracking against those benefits. We also plan to expand the task set by including multiple tasks within each gesture class. This could distinguish gesture-level effects from task-specific effects and provide stronger guidance for designing haptic feedback for VR interaction.

\section{Conclusion}

We examined how input and haptic feedback jointly shape gesture-based VR interaction through a within-subjects study in which participants completed grasp-, pinch-, and tap-like tasks using controller-based or hand-tracked input paired with one of three haptic conditions\textemdash{}vibration, impulse, or none. 
We implemented custom vibration and impulse feedback for both inputs to support a more controlled comparison in VR interaction. 
Our evaluation considered both objective task performance and subjective experience.

Our findings show that input, gesture, and haptic feedback jointly shape VR interaction. Controllers generally supported stronger objective performance in speed and contact control, while hand tracking was evaluated more positively for comfort, ease of use, and directness. Haptic feedback had limited effects on objective performance, but played a clearer role in subjective experience, where active feedback was rated more favourably than no feedback and was associated with contact confirmation, confidence, realism, and engagement.

These results suggest that haptic feedback should not be treated as a generic enhancement. Its value depends on the input used, the gesture performed, and the type of feedback delivered. By comparing matched haptic mechanisms across controller-based and hand tracking interaction, this work provides evidence that VR haptics should be designed and evaluated as interaction-specific cues rather than as interchangeable additions to an input device.

\bibliographystyle{abbrv}
\bibliography{references}

\end{document}